\documentclass[useAMS, usenatbib]{biom}
\usepackage{amsmath}
\usepackage{amsfonts}
\usepackage{amssymb}
\usepackage{graphicx}
\usepackage{float}
\usepackage{booktabs, tabularx}
\usepackage{makecell}
\usepackage{enumitem}
\usepackage{tcolorbox}
\usepackage{algorithm}
\usepackage{algpseudocode} 

\newlist{assump}{enumerate}{1}
\setlist[assump]{label=\textbf{A\arabic*.},ref=A\arabic*,
                 leftmargin=*,      
                 labelsep=0.6em,    
                 itemsep=0.25ex,    
                 topsep=0.5ex,      
                 align=left}        

\def\bSig\mathbf{\Sigma}

\newcommand{\bo}[1]{\mathbf{#1}}
\newcommand{\bs}[1]{\boldsymbol{#1}}
\newcommand{\sign}{\mathrm{sign}}
\newcommand{\abs}[1]{\lvert #1 \rvert}
\newcommand{\GIG}{\mathcal{GIG}}
\newcommand{\IG}{\mathcal{IG}}

\newcommand{\mat}[1]{\begin{bmatrix}#1\end{bmatrix}}

\newcommand{\diag}{\operatorname{Diag}}

\algrenewcommand\algorithmicrequire{\textbf{Require:}}
\algrenewcommand\algorithmicensure{\textbf{Ensure:}}
\algnewcommand\algorithmicinitialize{\textbf{Initialize:}}
\algrenewcommand\algorithmiccomment[1]{\hfill\(\triangleright\) #1}
\algnewcommand\Initialize{\item[\algorithmicinitialize]}

\title[Backward Bayesian Outcome Weighted Learning]{Backward Bayesian Outcome Weighted Learning}

\author{Emmanuel M. Rockwell$^{1,*}$\email{erockwell@unc.edu}, Michael R. Kosorok$^{1}$, and Nikki L. B. Freeman$^{2}$  \\
$^{1}$Department of Biostatistics, University of North Carolina at Chapel Hill, Chapel Hill, North Carolina, U.S.A. \\
$^{2}$Department of Biostatistics and Bioinformatics, Duke University, Durham, North Carolina, U.S.A.}

\begin{document}





\pagerange{} 
\volume{}
\pubyear{}
\artmonth{}


\doi{}


\label{firstpage}


\begin{abstract}
A central objective of precision medicine is learning optimal dynamic treatment regimes (DTRs) from data. Classification-based methods, like outcome weighted learning (OWL) for single-stage and backward OWL (BOWL) for multi-stage problems, leverage machine learning to directly learn optimal DTRs. However, these methods lack a natural way to quantify uncertainty in treatment decisions at the individual level. In this paper, we extend Bayesian OWL, a Bayesian reformulation of OWL, to the multi-stage setting. We call this method backward Bayesian outcome weighted learning (BBOWL). Like BOWL, our method directly learns an optimal DTR via backward induction, and unlike existing methods, our approach propagates uncertainty backward through the DTR learning process and provides uncertainty quantification of individualized treatment recommendations. We present a theoretical justification of BBOWL and verify its performance via a simulation study.
\end{abstract}

%



\maketitle


%

\section{Introduction}
\label{s:intro}

A central aim of precision medicine is to use adaptive, personalized decision-making to match patients with treatments that maximize clinical benefit \citep{Kosorok2021-qm}. Dynamic treatment regimes (DTRs) are a sequence of decision rules that formalize this approach by mapping a patient's history to treatment recommendations at key decision points \citep{Kosorok2021-qm, Chakraborty2013-yq}. 

Early work on DTRs established both causal inference and reinforcement learning frameworks for sequential decision making \citep{Murphy2003-sm, Schulte2014-zb}. In causal inference, structural equation models and marginal structural models formalize identification of optimal regimes under time-varying confounding \citep{Robins2008-wa, Orellana2010-mj}. Building on these methods, regression-based estimation strategies such as Q-learning and A-learning use conditional mean modeling to derive decision rules in a two-step fashion by first modeling expected outcomes given history and treatment and then using this conditional expectation to derive decision rules \citep{Chakraborty2013-yq, Schulte2014-zb}. These regression-based approaches are useful in many contexts, but can suffer when the conditional mean model is misspecified, leading to suboptimal treatment recommendations \citep{Murphy2005-xx}. 

Outcome weighted learning (OWL) introduces an alternative by formulating the goal of deriving an ITR as a weighted classification problem \citep{Zhao2012-yr}. In the single-stage setting, OWL recasts the objective of outcome maximization as the minimization of a weighted 0-1 loss function \citep{Hastie2009-lc}. Rather than modeling the outcome, OWL directly estimates the ITR by utilizing weighted support vector machines to classify assignments relative to a dividing hyperplane.

The multi-stage setting introduces an interdependence between treatment decisions at different decision points. The naive approach of estimating each stage's rule in isolation risks ignoring downstream effects such as interactions and delayed effects, ultimately producing suboptimal overall regimes \citep{Kosorok2021-qm}. Backward outcome weighted learning (BOWL) resolves this issue by applying a backward inductive framework, optimizing rules recursively from the final stage to the first to ensure the estimated regime maximizes the total cumulative outcome across all stages \citep{Zhao2015-ir, Bellman1957-fx}. At each stage, BOWL weights observations using estimated cumulative outcomes conditional on adherence to the previously estimated future stage rules, ensuring that earlier stage decisions are optimized with respect to the long-term expected outcome.

Bayesian outcome weighted learning complements OWL by embedding OWL's hinge-loss risk function in a Bayesian framework to enable uncertainty quantification \citep{Yazzourh2024-sx}. While OWL provides point estimates of individualized treatment rules, it lacks a natural mechanism for quantifying uncertainty in treatment recommendations. The Bayesian formulation of OWL is inspired by data augmentation for support vector machines, which expresses the pseudo-likelihood as a scale mixture of Gaussians \citep{Polson2011-bo}. This results in a pseudo-posterior, or Gibbs posterior, distribution from which samples can be drawn (e.g. via a Gibbs sampler), enabling posterior inference and the construction of credible intervals for uncertainty quantification \citep{Martin2022-vf}.

No existing method simultaneously accommodates both the backward recursive optimization required for multi-stage decision problems and the probabilistic framework needed to quantify uncertainty in individualized treatment rules. We propose backward Bayesian outcome weighted learning (BBOWL), which harmonizes the dynamic programming structure of BOWL with a Bayesian formulation of the outcome weighted learning objective. This combination enables uncertainty quantification for dynamic treatment regimes while retaining the advantages of classification-based methods. Furthermore, we develop an uncertainty propagation paradigm that uses posterior draws across stages to properly account for uncertainty and provide Gibbs sampling updates under normal, exponential power, and spike-and-slab priors. We evaluate BBOWL through simulation studies comparing BBOWL under various prior specifications with BOWL.

Section~\ref{s:methods} presents the BBOWL methodology and identification conditions, Section~\ref{s:est} details the estimation strategy across a variety of prior specifications, Section~\ref{s:simulation} reports simulation results, and Section~\ref{s:discussion} provides concluding remarks and directions for future work.

\section{Methods}
\label{s:methods}

\subsection{Setting and notation}
We consider $K$ decision stages with binary treatment actions $A_k\in \mathcal{A}_k = \{-1,1\}$ at stage $k\in\{1,\dots,K\}$. Let $X_1$ be baseline covariates and, for $k\ge 2$, let $X_k$ be information observed after $A_{k-1}$ and before $A_k$. Define the history
\[H_k = \left(\overline{X}_k, \overline{A}_{k-1}\right) = \left(X_1, A_1, \ldots, A_{k-1}, X_k,\right). \]
An overbar denotes the past and an underline represents the future such that $\overline{A}_k = (A_1, A_2, \ldots, A_{k})$ and $\underline{A}_k = (A_k, A_{k+1}, \ldots, A_K)$. Furthermore, note that $H_1 = X_1$ and $H_K = (\overline{X}_K, \overline{A}_{K-1})$. The primary outcome $Y \in \mathbb{R}^+$ is measured after stage $K$. The final vector of collected information is $(X_1, A_1, \ldots X_K, A_K, Y)$.

We observe $n$ independent and identically distributed trajectories $\{(X_{i,1}, A_{i,1}, \ldots, X_{i,K}, A_{i,K}, Y_i)\}_{i=1}^n$. Correspondingly, for each stage $k$, the sample-level quantities $A_k=(A_{1,k},\ldots,A_{n,k})^\top$, $X_k=(X_{1,k},\ldots,X_{n,k})^\top$, and $H_k=(H_{1,k},\ldots,H_{n,k})^\top$ collect the observations across all subjects. We use the subscript $i$ to refer to individual-level quantities and suppress it when referring to vectors over the sample. Write the stage-$k$ treatment probability as 
\[\rho_k(a_k\mid h_k)=P(A_k=a_k \mid H_k=h_k),\qquad a_k\in\{-1,1\}.\]

A dynamic treatment regime (DTR) is a sequence of decision rules $d=(d_1,\dots,d_K)\in\mathcal{D}$ with $d_k:\mathcal{H}_k\to\mathcal{A}_k$. Under a particular regime, the assigned treatment at stage $k$ is $A_k = d_k(H_k)$. Our goal is to estimate the optimal regime $d^{\text{opt}} = (d_1^{\text{opt}}, \ldots, d_K^{\text{opt}})$ that maximizes the expected outcome such that $E[Y^*(d)]\leq E[Y^*(d^{\text{opt}})]$ for all $d\in\mathcal{D}$.

\subsection{Identification assumptions}
Let $Y^*(\overline a_K)$ and $X_k^*(\overline a_{k-1})$ denote potential outcomes and covariate processes under treatment sequence $\overline a_K=(a_1,\dots,a_K)$. We assume:

\begin{enumerate}[label=\textbf{A\arabic*.},leftmargin=*]
\item (Consistency) When $\overline A_k = (a_1,\ldots, a_k)$ and $\overline A_{k-1} = (a_1,\ldots, a_{k-1})$ are observed, then $Y=Y^*(\overline A_K)$ and $X_k=X_k^*(\overline A_{k-1})$ with no interference and well-defined interventions.
\item (Positivity) For all $k$, $a_k$, and $H_k\in\mathcal{H}_k$, there exists $c>0$ such that $\rho_k(a_k\mid H_k)\ge c$ almost surely.
\item (Sequential randomization assumption) For all $\overline a_K$,
\[
A_k \perp\!\!\!\perp \big\{X_k^*(\overline a_{k-1}),\dots, X_K^*(\overline a_{K-1}), Y^*(\overline a_K)\big\}\;\big|\; H_k .
\]
\end{enumerate}

These assumptions parallel those introduced in the marginal structural modeling literature for causal inference with time-varying treatments \citep{Robins2008-wa, Orellana2010-mj}. Under these assumptions, the potential outcome under a dynamic treatment regime $d=(d_1,\dots,d_K)$ is defined as
\[Y^*(d) = Y^*\left(d_1(H_1),\,d_2(H_2),\,\dots,\,d_K(H_K)\right),\]
that is, the outcome that would be observed if, at each stage $k$, treatment $A_k$ were assigned according to the decision rule $d_k(H_k)$.

\subsection{Our approach}

Backward OWL proceeds by stepwise backward induction \citep{Zhao2015-ir}. Suppose optimal rules for future stages are known, denoted $\underline d_{k+1}^{\text{opt}}=(d_{k+1}^{\text{opt}},\ldots,d_K^{\text{opt}})$.
At stage $k$, the value of a candidate rule $d_k$ is
\begin{equation}
\label{eq:value}
V(d_k)\! =\! E\!\left[\frac{Y I\{\underline{A}_{k+1}\! =\! \underline{d}_{k+1}^{\text{opt}}(H_{k+1})\}}{\rho(\underline{a}_k|h_k)} I\{A_k=d_k(H_k)\}\right],
\end{equation}
where $\rho(\underline{a}_k|h_k) = \prod_{i=k}^K \rho(a_i\mid h_i)$ is the known or estimated treatment assignment probability at stage $k$. This factor normalizes by the probability of the observed treatment trajectory so that subjects with rarer treatment paths are given greater influence. The indicator term $I(\underline{A}_k = \underline{d}_k(H_k)) = \prod_{i=k}^K I(A_i = d_i(H_i))$ ensures that only individuals whose future treatment paths conform to the optimal rules at subsequent stages contribute to the current-stage estimation. 
The optimal rule maximizes \eqref{eq:value} and equivalently minimizes the weighted misclassification risk
\[d_k^{\text{opt}} = \arg\min_{d_k\in\mathcal{D}} E\left[\frac{Y I\{\underline{A}_{k+1}\! =\! \underline{d}_{k+1}^{\text{opt}}(H_{k+1})\}}{\rho(\underline{a}_k|h_k)} I\{A_{k}\neq d_k(H_{k})\}\right].\]

To simplify notation, at each stage $k$, denote the weight for observation $i$ as
\[w_{i,k} = \frac{y_i I\{\underline{A}_{i,k+1} = \underline{d}_{k+1}^{\text{opt}}(H_{i,k+1})\}}{\rho(\underline{a}_{i,k}|h_{i,k})}\]

Following OWL, we replace the 0-1 loss with a surrogate convex hinge loss to facilitate optimization \citep{Zhao2012-yr}. Furthermore, we can parameterize $d(x) = \sign\{f_k(H_k;\bs\beta_k)\}$ for some $f_k\in\mathcal{F}$. Restricting to a linear class $f_k(h)=h_k^\top\bs\beta_k$ yields the empirical objective function
\begin{equation}
\label{eq:loss}
    Q_k(\beta_k) = \sum_{i=1}^n w_{i,k}(1 - A_{i,k} H_{i,k}^\top\beta_{i,k})_+ + \sum_{j=1}^p p_\lambda (\abs{\beta_{j,k}})
\end{equation}
where $(z)_+ = \max(z,0)$ is the aforementioned hinge loss, $p_\lambda(\bs\beta_k)$ is a sparsity-inducing penalty function, and $\lambda$ is a tuning parameter. The penalty term promotes parsimony and stabilizes estimation in high-dimensional settings by shrinking small coefficients toward zero. In our Bayesian framework, these penalties correspond to distinct prior distributions on $\beta_k$, described in Section~\ref{s:priors}.

Utilizing Bayesian support vector machine results from \citet{Polson2011-bo}, the hinge loss can be expressed as a location-scale mixture of normals using latent variables. This creates a pseudo-likelihood (the empirical risk function), which, when combined with a prior, produces a pseudo-posterior, more commonly referred to as a Gibbs posterior, whose mode corresponds to the minimizer of equation (\ref{eq:loss}) \citep{Martin2022-vf}.
\begin{align*}
	p(\bs\beta_{k}\mid H_k)&\propto \exp(-Q_k(\bs\beta_k))\\
	&\propto \exp\left\{-\sum_{i=1}^n w_{i,k}(1 - A_{i,k} H_{i,k}^\top\beta_{i,k})_+\right\} \prod_{j=1}^p p(\beta_{j,k})\\
	&\propto C_k L(a_k|\bs\beta_k)p(\bs\beta_k)
\end{align*}
From here, \citet{Polson2011-bo} show that $L(a_k|\bs\beta_k)$ is a location-scale mixture of normals, and we can choose a suitable prior $p(\bs\beta_k)$ to find a solution of the $k$th stage \citep{Polson2011-bo}.

\subsection{Pseudo-likelihood construction}
\label{s:lik}

Let $u_{i,k} = 1-a_ih_i^\top\boldsymbol\beta_k$ for observation $i$ at stage $k$. Consider the identity
\[\int_0^\infty \phi(u\mid -\lambda, \lambda)d\lambda = e^{-2\max(u, 0)}\]
where $\phi(\;\cdot\mid m,v)$ denotes the normal probability distribution function with mean $m$ and variance $v$ \citep{Polson2011-bo, Andrews1974-il}. Then, the contribution of a single observation to the pseudo-likelihood for BBOWL can be expressed as

\begin{align*}
	L_i(a_i|\beta_k) &= \exp\left\{-2w_{i,k}\max(1-a_ih_i^\top\beta_{k}, 0)\right\}\\
	&= \int_0^\infty \frac{1}{2\pi\lambda_i}\exp\left\{-\frac{\left(\lambda_i + w_{i,k}u_{i,k}\right)^2}{2\lambda_i}\right\}d\lambda_i
\end{align*}



\subsection{Prior specification}
\label{s:priors}
The choice of prior distribution for the DTR parameter $\bs\beta_k$ plays a central role in backward Bayesian outcome weighted learning. Using both the Bayesian OWL framework of \citet{Yazzourh2024-sx} and the Bayesian SVM representation of \citet{Polson2011-bo}, we consider three families of priors that correspond to different regularization techniques. 

\subsubsection{Normal prior distribution for $\beta_k$}
\label{s:normspec}

A natural baseline is to assume a Gaussian prior on the DTR parameters $\bs\beta_k \sim MVN(\bs\mu_k, \Sigma_k)$ where $\bs\mu_k = (\mu_{k,1},\ldots, \mu_{k,p})$ and $\Sigma_k = \diag(\sigma_1^2,\ldots, \sigma_p^2)$. With this normal prior, we can write the pseudo-posterior distribution as
\begin{align*}
	&p(\bs\beta_k, \boldsymbol\lambda \mid \bo h, \bo y, \bo a, \bs\mu_k, \Sigma_k) \\
    &\propto  \prod_{i=1}^n \frac{1}{\sqrt{2\pi\lambda_i}} \exp\left\{ - \frac{(\lambda_i + w_{i,k}u_{i,k})^2}{2\lambda_i} \right\}\\
    &\qquad \times\prod_{j=1}^p \frac{1}{\sqrt{2\pi\sigma_j^2}}\exp\left\{-\frac{(\beta_{j,k} - \mu_j)^2}{2\sigma_j^2}\right\}
\end{align*}
where $\bs\lambda = (\lambda_1,\ldots, \lambda_n)^\top$, $\bo h = (h_1,\ldots, h_n)$, $\bo y = (y_1,\ldots, y_n)$, and $\bo a = (a_1,\ldots, a_n)$. With regard to variable selection, the normal prior is the Bayesian analog of a ridge-type penalty \citep{Goldstein1974-ql}.

\subsubsection{Exponential power prior distribution for $\beta_k$}
\label{s:exppowspec}

To encourage shrinkage and sparse rules, one may instead use the exponential power distribution \citep{Gomez-Sanchez-Manzano2008-ot, Gold2005-me}. Let $\bs\omega = (\omega_1,\ldots, \omega_p)$ and $\omega_j >0$. \citet{Polson2011-bo} formulate the double exponential prior regularization penalty as
\[L_i(y_i|\bs\beta) = \int_0^\infty \phi(\beta_j|0, \nu_k^2\omega_j\sigma_j^2)p(\omega_j | \alpha) d\omega_j\]
where $p(\omega_j|\alpha) \propto \omega_j^{-3/2}St_{\frac{\alpha}{2}}^+(\omega_j^{-1})$ and $St_\frac{\alpha}{2}^+$ is the density function of a positive stable random variable of index $\alpha/2$ \citep{Polson2011-bo, West1987-wc}.  Under this prior distribution specification, the corresponding pseudo-posterior distribution is
\begin{align*}
	&p(\bs\beta_k, \boldsymbol\lambda \mid \bo h, \bo y, \bo a, \nu_k, \bs\omega) \\
    &\propto  \prod_{i=1}^n \frac{1}{\sqrt{2\pi\lambda_i}} \exp\left\{ - \frac{(\lambda_i + w_{i,k}u_{i,k})^2}{2\lambda_i} \right\}\\
	&\qquad \times \abs{\Omega}^{-\frac{1}{2}}\exp\left\{-\frac{1}{2\nu_k^2}\sum_{j=1}^p \frac{\beta_{j,k}^2}{\sigma_j^2\omega_j}\right\}\times \prod_{j=1}^p p(\omega_j|\alpha)
\end{align*}

where $\Omega \equiv \diag(\omega_1,\ldots, \omega_p)$. When $\alpha = 1$, we have $p(\omega_j|\alpha)\sim Exp(2)$ which recovers the Bayesian lasso penalty structure \citep{Polson2011-bo, Hans2009-np}. The case when $\alpha = 2$ corresponds to ridge regression \citep{Goldstein1974-ql}. \citet{West1987-wc} established this result for $\alpha\in[1,2]$, and \citet{Gomez-Sanchez-Manzano2008-ot} extended it to $\alpha\in(0,1]$, thereby encompassing the ``bridge estimator" family of \citet{Huang2008-qm}. This prior unifies the ridge ($\alpha=2$) and lasso ($\alpha=1$) penalties as special cases, while allowing intermediate values of $\alpha$ to yield heavier tails and stronger concentration near zero, promoting adaptive sparsity. Our estimation strategy and simulation will focus on the case where $\alpha = 1$.

\subsubsection{Spike-and-slab prior distribution for $\beta_k$}
\label{s:ssspec}

Finally, to encourage sparsity while retaining conjugate Gaussian updates, we adopt a normal ``continuous spike-and-slab" prior for the ITR coefficients. For each $k$, introduce 
\[\gamma_j\sim Bernoulli(\pi)\;\; \textrm{and}\;\; \beta_j|x_j\sim N(0,\tau^2_{\gamma_j}\sigma_j^2)\;\; \textrm{with}\;\; \tau^2_0 \ll \tau_1^2.\] 

In other words, $N(0,\tau_0^2\sigma_j^2)$ corresponds to the ``spike" at 0 and $N(0, \tau_1^2\sigma_j^2)$ corresponds with the ``slab" component. The continuous mixture leaves the linear predictor as $h_k^\top\bs\beta_k$ and expresses sparsity through the prior precision $\Sigma_\gamma^{-1}$ \citep{George1993-kj, Mitchell1988-op}. Under this prior distribution specification, the corresponding pseudo-posterior distribution is
\begin{align*}
	&p(\bs\beta_k, \boldsymbol\lambda \mid \bo h, \bo y, \bo a, \bs\mu_k, \Sigma_k)\\
    &\propto  \prod_{i=1}^n \frac{1}{\sqrt{2\pi\lambda_i}} \exp\left\{ - \frac{(\lambda_i + w_{i,k}u_{i,k})^2}{2\lambda_i} \right\}\\
	&\qquad \times \prod_{j=1}^p \left[\pi\phi\left(\beta_{j,k}| 0, \tau_1^2\sigma_j^2\right)\right]^{\gamma_j}\left[(1-\pi)\phi\left(\beta_{j,k}|0, \tau_0^2\sigma_j^2 \right)\right]^{1-\gamma_j}
\end{align*}
where $\phi(\;\cdot \mid m, v)$ is a normal probability distribution function with mean $m$ and variance $v$. 

\section{Estimation}
\label{s:est}

The value of a Bayesian formulation in part lies in the ability to draw from the pseudo-posterior distribution. We will be utilizing the Gibbs sampling algorithm introduced by \citet{Polson2011-bo} to enable sampling from the pseudo-posterior. While this may be more computationally intensive than the expectation-maximization (EM) alternative given by \citet{Polson2011-bo}, the Gibbs sampling approach enables uncertainty quantification \citep{Polson2011-bo, Arjas2010-an}. 

\subsection{Estimation with the Gaussian prior}
The pseudo-posterior under a normal prior for $\bs\beta_k$ depends on two unknown parameters: the regression coefficients $\bs\beta_k$ and the latent augmentation variables $\bs\lambda$. To generate samples from this posterior, we employ a Gibbs sampling strategy, alternately updating each parameter block given the most recent values of the others. In this section, we provide a concise overview of the derivation, with full details provided in the Web Appendix. Specifically, the conditional distribution of $\bs\lambda|\bs\beta_k, \bo h, \bo y, \bo a$ can be expressed as

\begin{align*}
	p(\bs\lambda \mid \bs\beta_k, \bo h, \bo y, \bo a)
	&\propto \prod_{i=1}^n \lambda^{-\frac{1}{2}}\exp\left\{-\frac{\lambda_i}{2}\right\} \exp\left\{-\frac{({w_{i,k}}^2u_{i,k}^2)}{2\lambda_i}\right\}
\end{align*}
According to \citet{Devroye1986-kx}, a random variable has a generalized inverse Gaussian distribution $\GIG(\gamma, \psi, \chi)$ if its density function is $p(x|\gamma, \psi, \chi) = C(\gamma, \psi, \chi)x^{\gamma - 1}\exp\{-\frac{1}{2}(\frac{\chi}{x}+\psi x)\}$ where $C(\gamma, \psi, \chi)$ is a normalization constant \citep{Devroye1986-kx}. Therefore, we can conclude that
\begin{align*}
    \lambda_i \mid \bs\beta_k, h, y, a &\sim \GIG\left(\frac{1}{2}, 1, {w_{i,k}}^2u_{i,k}^2\right)
\end{align*}
Note that if $X\sim \GIG(1/2, \alpha, \beta_k)$, then $X^{-1}\sim \GIG(-1/2, \beta_k, \alpha)$. Furthermore, the inverse Gaussian distribution $\IG(\mu, \alpha)$ is a special case of the generalized inverse Gaussian distribution with $p = -1/2$ where $\beta_k = \frac{\alpha}{\mu^2}$. Thus, $X^{-1}\sim \IG(\mu, \alpha)$. It follows that
\[\lambda_i^{-1} \mid \bs\beta_k, h, y, a \sim \IG\left(\frac{1}{{w_{i,k}}u_{i,k}} , 1\right)\]

The conditional distribution of $\bs\beta_k \mid \bs\lambda, \bo x, \bo y, \bo a$ is given in Section~\ref{s:normspec}. Define
\[\bo H \equiv \mat{a_1h_{1,1} & \cdots & a_1h_{1,p} \\ \vdots & \ddots & \vdots \\ a_{n}h_{n, 1} & \cdots & a_{n}h_{n, p}}_{(n\times p)},\quad \bo W = \mat{1+\frac{w_1}{\lambda_1} \\ \vdots \\ 1 + \frac{w_n}{\lambda_{n}}}_{(n\times 1)}\]
\[\bs\Lambda = \diag(\lambda_1, \ldots, \lambda_{n}),\quad\textrm{and}\quad\bo R = \diag(w_1, \cdots, w_{n})_{(n\times n)}.\]
Then, this conditional distribution can be expressed as
\begin{align*}
	P(\boldsymbol\beta_k \mid \boldsymbol\lambda, \bo x, \bo y, \bo a) \exp\left\{-\frac{1}{2}(\bs\beta_k - B_1b_1)^\top B^{-1}(\bs\beta_k - B_1b_1)\right\}
\end{align*} 
where
\[B_1 = (\bo H^\top \bo R\bs\Lambda^{-1}\bo R\bo H + \bs\Sigma^{-1})^{-1}\;\;\textrm{and}\;\; b_1 = \bo H^\top \bo R \bo W + \bs\Sigma_k^{-1}\bs\mu_k\]
Full details are provided in the Web Appendix. It follows that the conditional posterior is $\bs\beta_k | \bs\lambda, X \sim N(B_1b_1, B_1)$. Details of the Gibbs sampling scheme are given in Algorithm~\ref{alg:gibbs-normal}.

\begin{algorithm}[t]
\caption{Gibbs sampling for BBOWL with normal priors at stage $k$}
\label{alg:gibbs-normal}
\begin{algorithmic}[1] 
\Require Observed data $\{(h_{i,k},a_{i,k},y_i)\}_{i=1}^n$ and treatment weights $\{w_{i,k}\}_{i=1}^n$.

\Initialize Set initial values $(\bs\beta_k^{(0)},\bs\lambda^{(0)})$ and hyperparameters $(\bs\mu_k,\Sigma_k)$.

\State Draw $\bs\beta_k^{(g+1)}|\bs\lambda^{(g)}, \bo h, \bo y, \bo a \sim N(B_1^{(g)}b_1^{(g)}, B_1^{(g)})$
\State Draw $\bs\lambda^{-1(g+1)}|\bs\beta_k^{(g)}, \bo h, \bo y, \bo a$ where
    \[\lambda_i^{-1} | \bs\beta_k, \bo h, \bo y, \bo a \sim \IG\left(\frac{1}{{w_{i,k}}u_{i,k}} , 1\right)\]
\State Repeat steps 1 and 2 until chains converge.
\end{algorithmic}
\end{algorithm}

\subsection{Estimation with the exponential power prior}

When an exponential power prior is specified for $\bs\beta_k$, the resulting pseudo-posterior distribution involves estimating three unknown parameters: the regression coefficients $\bs\beta_k$, the latent augmentation variables $\bs\lambda$, and the local scale parameters $\bs\omega$. To explore this posterior, we develop a Gibbs sampling scheme in which each parameter block is updated sequentially given the most current values of the others. A concise summary of the full conditional distributions is given below, with full derivations given in the Web Appendix.

The conditional distribution $\bs\lambda \mid \bs\beta_k, \bo x, \bo y, \bo a$ is identical to that under the normal prior, since the augmentation of the hinge loss given by \citet{Polson2011-bo} does not depend on the choice of prior for $\bs\beta_k$. The conditional distribution of $\bs\beta_k \mid \bs\lambda, \bs\omega, \bo x, \bo y, \bo a$ is given in Section~\ref{s:exppowspec}. This conditional distribution can be expressed as

\begin{align*}
	p(\bs\beta_k | \bs\lambda, \bs\omega, \bo x, \bo y, \bo a) \propto \exp\left\{-\frac{1}{2}(\bs\beta - B_2b_2)^\top B_2^{-1}(\bs\beta - B_2b_2)\right\}
\end{align*}
where
\[
B_2 = (\bo H^\top \bo R\bs\Lambda^{-1}\bo R\bo H  + \nu_k^{-2}\bs\Omega^{-1}\bs\Sigma_k^{-1})^{-1}\;\;\textrm{and}\;\; b_2 = \bo H^\top \bo R \bo W\]

It follows that the conditional posterior is $\bs\beta_k | \bs\lambda, \bo h, \bo y, \bo a \sim N(B_2b_2, B_2)$. Finally, Corollary 3 of \citeauthor{Polson2011-bo} states that, for $\alpha = 1$, the full conditional distribution of $\omega$ is 
\[p(\omega_j^{-1}|\bs\beta_k, \nu_k) \sim IG\left(\frac{\nu_k\sigma_j}{\abs{\beta_{j,k}}}, 1\right)\]
Combining these three conditional distributions, we arrive at the Gibbs sampling scheme given by Algorithm~\ref{alg:gibbs-exppower}.

\begin{algorithm}[t]
\caption{Gibbs sampling for BBOWL with an exponential power prior at stage $k$}
\label{alg:gibbs-exppower}
\begin{algorithmic}[1] 
\Require Observed data $\{(h_{i,k},a_{i,k},y_i)\}_{i=1}^n$ and treatment weights $\{w_{i,k}\}_{i=1}^n$.

\Initialize Set initial values $(\bs\beta_k^{(0)},\bs\lambda^{(0)},\bs\omega^{(0)})$ and hyperparameters $(\nu_k, \Sigma_k)$.

\State Draw $\bs\beta_k^{(g+1)}| \nu_k, \bs\lambda^{(g)}, \bs\omega^{(g)}, \bo h, \bo y, \bo a \sim N(B_2^{(g)}b_2^{(g)}, B_2^{(g)})$
\State Draw $\bs\lambda^{-1(g+1)}|\bs\beta_k^{(g)}, \bo h, \bo y, \bo a$ where
    \[\lambda_i^{-1} | \bs\beta_k, \bo h, \bo y, \bo a \sim IG\left(\frac{1}{{w_{i,k}}u_{i,k}} , 1\right)\]
\State Draw $\omega_j^{-1(g+1)} | \beta_{j,k}^{(g+1)}, \nu_k \sim \IG(\nu_k\sigma_j\abs{\beta_{j,k}}^{-1}, 1)$
\State Repeat steps 1, 2, and 3 until chains converge.
\end{algorithmic}
\end{algorithm}

\subsection{Estimation with the spike-and-slab prior}
We will estimate three unknown parameters, $\bs\beta_k, \bs\lambda$ and $\bs\gamma_k$ when specifying a spike-and-slab prior on the pseudo-likelihood. The conditional distribution of $\lambda_i | \bs\beta_k, \bo h, \bo y, \bo a$ is the same as in the previous derivations.

We consider a continuous spike-and-slab prior on the stage-$k$ decision coefficients $\bs\beta_k=(\beta_{i,k},\ldots,\beta_{p,k})^\top$ such that
\[\gamma_j \sim \mathrm{Bernoulli}(\pi),\;\; \beta_{j,k}\mid \gamma_j \sim \mathcal N\big(0,\;\tau_{\gamma_j}^2 \sigma_j^2\big)\;\;\textrm{where}\;\; 
\tau_0^2\ll\tau_1^2.\]
Note that $\gamma_j\in\{0,1\}$ selects the spike ($\gamma_j=0$) or slab ($\gamma_j=1$) component. Define $D_\gamma^{-1} = \diag(\tau^{-2}_{\gamma_j}\sigma_1^{-2}, \ldots, \tau^{-2}_{\gamma_j}\sigma_p^{-2})$. The conditional distribution of $\bs\beta_k|\bs\lambda, \bs\gamma, \bo h, \bo y, \bo a$ is given in Section~\ref{s:ssspec}. This conditional distribution can be expressed as

\begin{align*}
	p(\bs\beta_k | \bs\lambda, \bs\omega, \bo h, \bo y, \bo a)& \propto \exp\left\{-\frac{1}{2}(\bs\beta_k - B_\gamma b_\gamma)^\top B_\gamma^{-1}(\bs\beta_k - B_\gamma b_\gamma)\right\}
\end{align*}
where
\[B_\gamma = (\bo H^\top \bo R\bs\Lambda^{-1}\bo R\bo H + D_\gamma^{-1})^{-1}\;\;\textrm{and}\;\;b_\gamma = \bo H^\top \bo R \bo W \]

Finally, because we use a continuous spike, the conditional of $\gamma$ updates by a Bernoulli draw with success probability
\[\Pr(\gamma_j=1\mid \beta_{k,j}) = \frac{\pi \phi \big(\beta_{k,j}| 0,\tau_1^2\sigma_j^2\big)}{\pi \phi\big(\beta_{k,j}| 0,\tau_1^2\sigma_j^2\big) + (1-\pi)\phi\big(\beta_{k,j}|0,\tau_0^2\sigma_j^2\big)}.\]
where $\pi$ is the prior on $\gamma$. As in the normal and exponential power cases, the stage-$k$ updates use weights $w_{i,k}$ that depend on posterior draws from stage $k+1$, thereby propagating uncertainty backward. The spike-and-slab prior adds model selection uncertainty at each stage via $\gamma$, which can be summarized by posterior inclusion probabilities (PIP) to quantify variable importance in each rule. In earlier stages, where uncertainty about future decisions is largest, the spike allows aggressive shrinkage, while the slab preserves signals that remain predictive after accounting for downstream rules. The resulting Gibbs sampling scheme is provided in Algorithm~\ref{alg:gibbs-spikeslab}.

\begin{algorithm}[t]
\caption{Gibbs sampling for BBOWL with a spike-and-slab prior at stage $k$}
\label{alg:gibbs-spikeslab}
\begin{algorithmic}[1] 
\Require Observed data $\{(h_{i,k},a_{i,k},y_i)\}_{i=1}^n$ and treatment weights $\{w_{i,k}\}_{i=1}^n$.

\Initialize Set initial values $(\bs\beta_k^{(0)},\bs\lambda^{(0)},\bs\gamma^{(0)})$ and hyperparameters $(\tau_0^2, \tau_1^2, \pi, \Sigma_k)$.

\State Draw $\beta_k^{(g+1)}| \bs\lambda^{(g)}, \bs\gamma^{(g)}, \bo h, \bo y, \bo a \sim N(B_\gamma^{(g)}b_\gamma^{(g)}, B_\gamma^{(g)})$
\State Draw $\bs\lambda^{-1(g+1)}|\bs\beta^{(g)}, \bo h, \bo y, \bo a$ where
    \[\lambda_i^{-1} | \bs\beta_k, \bs\gamma, \bo h, \bo y, \bo a  \sim \IG\left(\frac{1}{{w_{i,k}}u_{i,k}} , 1\right)\]
\State For each $j$ compute 
\[p_j = \frac{\pi \phi \big(\beta_{j,k}\mid 0,\tau_1^2\sigma_j^2\big)}{\pi \phi\big(\beta_{j,k}\mid 0,\tau_1^2\sigma_j^2\big) + (1-\pi)\phi\big(\beta_{j,k}\mid 0,\tau_0^2\sigma_j^2\big)}\] 
and draw $\gamma_j^{(g+1)} \sim \mathrm{Bernoulli}(p_j)$.
\State Repeat steps 1, 2, and 3 until chains converge.
\end{algorithmic}
\end{algorithm}

\subsection{Uncertainty propagation across stages}
\label{s:up}

While we have discussed the sampling algorithm for $\bs\beta_k$ within a specific stage $k$, we have yet to discuss how to carry information across stages. This is a relevant concern because estimation of $\bs\beta_k$ requires conditioning on the posterior of $\bs\beta_{k+1}$. Therefore, a key design choice concerns how to summarize this posterior distribution when carrying information backward.

For computation, one option is a deterministic plug-in update using the posterior mean $\hat{\bs\beta}_{k+1} = E(\bs\beta_{k+1}\mid H_{k+1})$. While this approach may reduce misclassification error, the estimate treats $\bs\beta_k$ as known by ignoring variability in the previous stage estimation and will therefore yield biased (typically downward) standard errors \citep{Murphy1985-yz}. Because variability in $\bs\beta_{k+1}$ is not propagated to earlier stages, posterior averaging produces anti-conservative inference.

BBOWL instead samples $\bs\beta_{k+1}$ directly from its posterior distribution as
\[\hat{\bs\beta}_{k+1} = \bs\beta_{k+1}\sim p(\bs{\beta}_{k+1}\mid H_{k+1})\]
and conditions on this draw when forming $w_{i,k}$ and updating $\bs\beta_k$. This strategy preserves the correct posterior dependence structure across stages and aligns with valid value inference recommendations for data-driven regimes \citep{Chakraborty2014-nb}. In practice, this amounts to a stage-wise Gibbs procedure in which posterior draws from later stages are reused as inputs to earlier stages, ensuring accurate uncertainty quantification for the full dynamic regime. Simulation results reported in Section~\ref{s:simulation} support this scheme.

\subsection{Prediction and decision uncertainty}
\label{s:uq}

A key advantage of this Bayesian formulation is the ability to quantify uncertainty in treatment recommendations at the individual level. Extending the framework of \citet{Yazzourh2024-sx}, let $\Theta_k = \{\bs\beta_k, \bs\lambda_k\}$ denote the stage-$k$ parameters and $\tilde a$ denote the recommended treatment for a new patient with features $\tilde h$. Then, it follows that
\begin{multline}
    p(\tilde a | \tilde h, H, y, a) = \int_{\Theta_k} p(\tilde a = 1 | \tilde h, H, y, a, \bs\beta_k, \bs\lambda_k)\\ \times p(\bs\beta_k, \bs\lambda_k | h, y, a) d\theta
\end{multline}

We adopt a probit link $p(a = 1 | h) = \Phi(h^\top\beta)$ for the stage-$k$ decision rule. Here, $\Phi$ denotes the cumulative distribution function of the standard normal distribution. Then, it follows that
\[p(\tilde a | \tilde h, H, \bo y, \bo a) = \int_{\Theta_k} \Phi(\tilde h^\top \bs\beta_k)p(\bs\beta_k, \bs\lambda_k | \bo x, \bo y, \bo a) d\theta\]
which is readily approximated by posterior samples of $\bs\beta_k$. The resulting posterior distribution of the score $\tilde h^\top\bs\beta_k$ provides both a point prediction via its sign and a measure of decision uncertainty via its dispersion around zero.

\section{Simulation study}
\label{s:simulation}

We conducted simulation studies to evaluate the performance of BBOWL in a three-stage treatment setting. The primary goals were: (i) to assess classification accuracy of BBOWL relative to BOWL under different prior specifications, (ii) to examine how performance varies with sample size and outcome structure, and (iii) to evaluate whether uncertainty propagation hinders predictive performance.

For all three criteria, classification performance is measured by the stage-wise misclassification rate, defined as the proportion of test subjects for whom the estimated treatment recommendation $d_k(h_{i,k})=\sign(h_{i,k}^\top\hat{\bs\beta}_k)$ differs from the oracle rule $d_k^{\mathrm{opt}}(h_{i,k})=\sign(h_{i,k}^\top\bs\beta_k^{\mathrm{true}})$ \citep{Schulte2014-zb}. 

\begin{figure*}
  \centerline{\includegraphics[width=\linewidth]{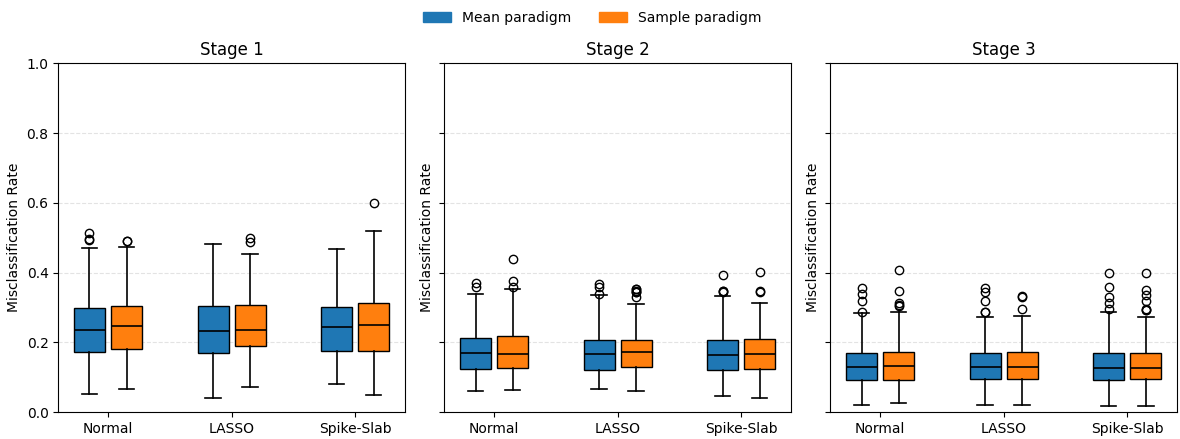}}
  \caption{Comparison of misclassification rates when propagating posterior means versus a single posterior draw. Results are shown for modeling a terminal outcome with N=1000 and 200 replicates.}
  \label{f:mean-vs-sample}
\end{figure*}

The uncertainty propagation scheme assessed by criterion (iii) is discussed in Section~\ref{s:up}. To evaluate the tradeoff in misclassification error, we compared the two strategies empirically across prior specifications. Figure~\ref{f:mean-vs-sample} reports misclassification rates across all three stages and prior specifications when substituting posterior means and propagating single posterior draws. The results indicate that both strategies yield similar performance in terms of misclassification. Importantly, however, only the sampling approach enables uncertainty quantification at the individual level, providing a more reliable reflection of posterior variability while maintaining comparable predictive accuracy.

\subsection{Design}
We evaluated the performance of the BBOWL in a three-stage treatment setting under the normal, exponential power, and spike-and-slab priors. At each stage $k\in\{1,2,3\}$, the treatment $A_k\in \{-1,1\}$ is assigned with equal probability independently of covariates. Consequently, the treatment propensity is constant across subjects and stages, $\rho_k(a_k\mid H_k)=0.5$ for all $k$ and $a_k$. 

Each simulated dataset contains both prognostic and prescriptive covariates. Prognostic covariates are $X=(1_p,X_1)$, with $X_1\sim \text{Bernoulli}(0.5)$. At each stage $k$, we generated prescriptive covariates $W_k=(1,W_{k1},\ldots,W_{k5})$ with $W_{kj}\sim \text{Uniform}(0,1)$ independently. Two outcome-generating mechanisms, one with a single terminal outcome and one with intermediate outcomes recorded at each stage, were considered:
\begin{align*}
    &\textrm{Terminal Outcome:}\\ &\qquad Y^* = X\eta + A_1W_1^\top\beta_1 + A_2W_2^\top\beta_2 + A_3W_3^\top\beta_3 + \epsilon\\ 
    &\textrm{Intermediate Outcome:}\\
    &\qquad  Y_k^* = X\eta + A_k W_k^\top \beta_k + \epsilon_k,\quad k=1,2,3,
\end{align*}
with $\epsilon_k, \epsilon\sim N(0,\sigma^2)$ with $\sigma^2 = 0.05$. To ensure strictly positive outcomes, we apply a global shift $c = \max\{0, -\min(Y^*, Y_1^*, Y_2^*, Y_3^*)\} + 0.1$ and set $Y = Y^* + c$ and $Y_k = Y_k^* + c$. The true coefficients are
\begin{align*}
    \eta &= (1.0, -0.5)\\
    \beta_1 &= (1.5, -3.5, 2.0, 0, 0, 0)\\
    \beta_2 &= (1.0, 0, 3.0, 0, -4.0, 0)\\
    \beta_3 &= (-2.5, 0, 0, 3.5, 0, 2.0)
\end{align*}
The corresponding stage-$k$ optimal treatment is the linear decision rule $A_k^* = \sign(W_k^\top\bs\beta_k)$. Posterior sampling is via Gibbs updates tailored to each prior. Posterior inference was obtained via Gibbs sampling with 1000 iterations per dataset, discarding the first 50 as burn-in.  

We vary the individual dataset sample sizes over $N\in\{400, 600, 800, 1000\}$. For each $N$, we generate 200 replicate datasets. Within each replicate, we split into 70\% training and 30\% test sets using fixed random seeds to ensure reproducibility and balance across iterations. Separate experiments were run under the terminal-outcome and intermediate-outcome data-generating mechanisms. In the terminal outcome setting, we only use $Y$ when estimating each stage rule and in the intermediate outcome setting, we use $Y_k$ when estimating the stage-$k$ rule. 

\begin{figure*}
  \centering
  \includegraphics[width=\linewidth]{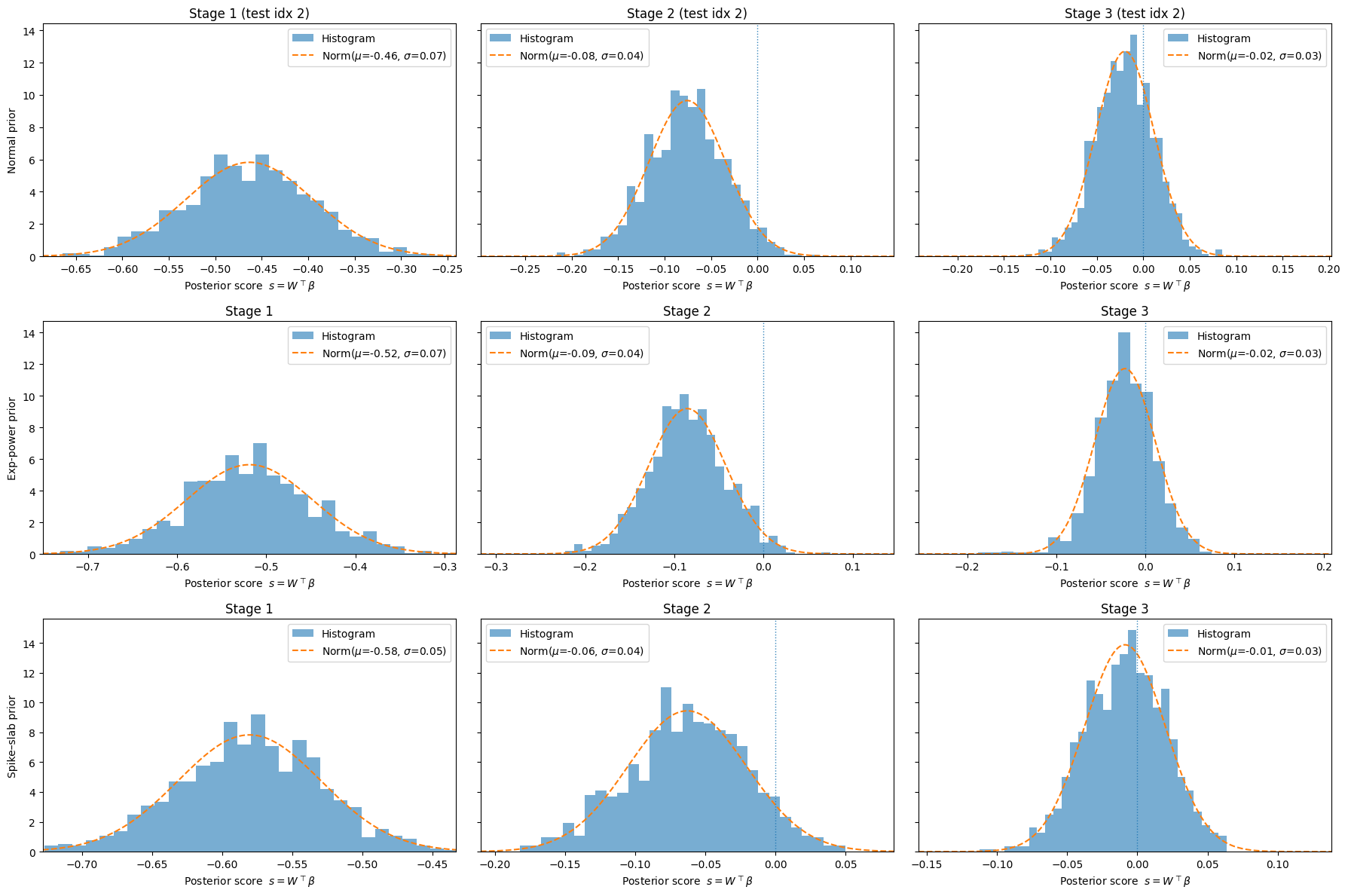} 
  \caption{Treatment recommendation uncertainty quantification for a single test set subject across all stages and prior specifications.}
  \label{fig:uncertainty-quant}
\end{figure*}

Figure~\ref{fig:uncertainty-quant} illustrates the posterior distribution of the individual-level decision-score $f_k(h_k;\bs\beta_k) = h_k^\top\bs\beta_k$ for a single test set subject across all three stages and prior specifications. As mentioned earlier, the sign of this score function is the predicted optimal treatment assignment. Therefore, one can easily assess decision uncertainty via the proximity of the distribution to zero. In fact, the probability of either treatment being optimal is the proportion of the distribution that lies on each respective size of zero. Furthermore, as the recursion proceeds backward, the posterior distributions become progressively wider, reflecting the accumulation of uncertainty propagated from future to earlier decisions. When the distribution of $h_k\top\bs\beta_k$ is concentrated near 0, the sign of the score, and hence the optimal treatment recommendation $d_k(h_k) = \sign\{h_k^\top\bs\beta_k\}$, becomes unstable across posterior draws. In such cases, the probability of recommending treatment $A_k = 1$ or $A_k = -1$ is similar, indicating that the individual lies near the decision boundary where treatment benefit is ambiguous and recommendation reliability is low.

\subsection{Classification performance}
Performance was assessed by the misclassification rate, defined as the proportion of test subjects with treatment recommendations that differed from the true optimal rule at each stage. We report results separately by stage, sample size, and prior.

\begin{figure*}
  \centering
  \includegraphics[width=\linewidth]{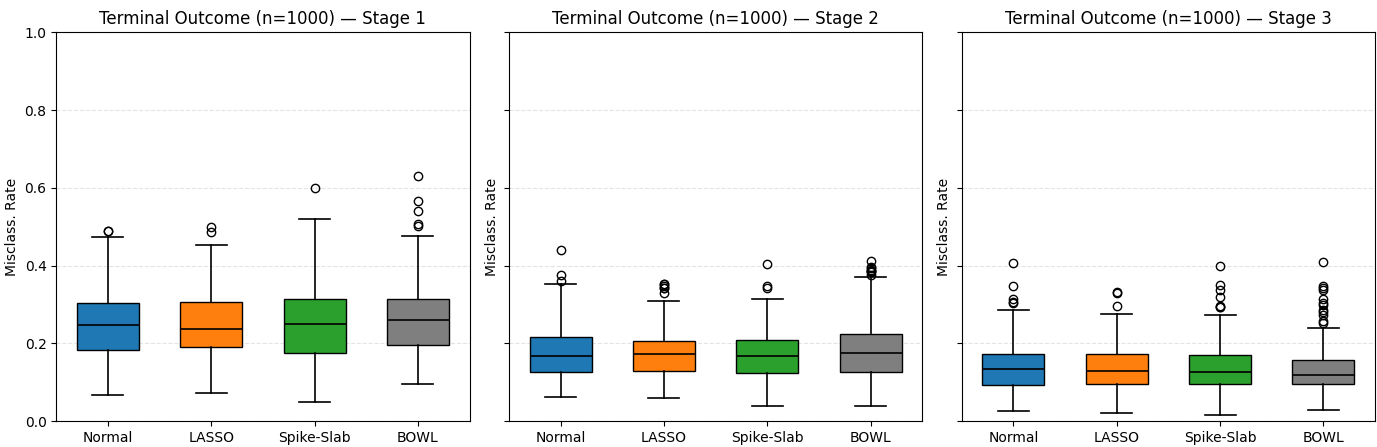} 
  \caption{BBOWL misclassification rates for all three prior specifications compared to BOWL. Reported for the $n=1000$ simulation scenario.}
  \label{fig:misclass-terminal}
\end{figure*}

Table~\ref{tab:misclass_terminal} and Figure~\ref{fig:misclass-terminal} summarize the results. Across all scenarios, misclassification rates decreased monotonically with increasing $N$, illustrating the improved reliability of both BOWL and BBOWL as sample size increases.  

\begin{table*}[t]
\centering
\footnotesize
\setlength{\tabcolsep}{3pt}       
\renewcommand{\arraystretch}{1.1} 
\begin{tabularx}{\textwidth}{r*{12}{>{\centering\arraybackslash}X}}
\toprule
 & \multicolumn{4}{c}{Stage 1} & \multicolumn{4}{c}{Stage 2} & \multicolumn{4}{c}{Stage 3} \\
\cmidrule(lr){2-5}\cmidrule(lr){6-9}\cmidrule(lr){10-13}
\(n\)
  & BOWL & \multicolumn{3}{c}{BBOWL Priors}
  & BOWL & \multicolumn{3}{c}{BBOWL Priors}
  & BOWL & \multicolumn{3}{c}{BBOWL Priors} \\
\cmidrule(lr){3-5}\cmidrule(lr){7-9}\cmidrule(lr){11-13}
 &  & Normal & \makecell{Exp.\\Power} & \makecell{Spike-\\and-slab}
 &  & Normal & \makecell{Exp.\\Power} & \makecell{Spike-\\and-slab}
 &  & Normal & \makecell{Exp.\\Power} & \makecell{Spike-\\and-slab} \\
\midrule
400  & 0.31 & 0.32 & 0.32 & 0.31 & 0.23 & 0.22 & 0.22 & 0.23 & 0.19 & 0.20 & 0.20 & 0.20 \\
600  & 0.29 & 0.28 & 0.28 & 0.28 & 0.22 & 0.20 & 0.20 & 0.20 & 0.17 & 0.17 & 0.17 & 0.17 \\
800  & 0.27 & 0.26 & 0.26 & 0.26 & 0.21 & 0.19 & 0.19 & 0.19 & 0.14 & 0.16 & 0.16 & 0.16 \\
1000 & 0.26 & 0.25 & 0.25 & 0.25 & 0.19 & 0.18 & 0.17 & 0.17 & 0.13 & 0.14 & 0.14 & 0.14 \\
\bottomrule
\end{tabularx}
\caption{Misclassification rates for each stage and sample size under BOWL and Bayesian variants (terminal outcome scenario).}
\label{tab:misclass_terminal}
\end{table*}

Under the terminal outcome paradigm (Table~\ref{tab:misclass_terminal}), BBOWL achieved performance nearly identical to BOWL at all stages and sample sizes, regardless of prior specification. Stage 3 generally exhibited the lowest misclassification rates, reflecting the accumulation of information closer to the final outcome. Figure~\ref{fig:misclass-terminal} visualizes the similarity of misclassification rates across methods as $N$ increases.

Overall, the simulation results demonstrate that: (i) BBOWL performs comparably to BOWL in terms of classification accuracy, with no systematic loss from propagating posterior uncertainty; (ii) misclassification rates lower steadily with larger sample sizes; and (iii) the choice among normal, exponential power, or spike-and-slab priors has little effect on classification accuracy. These findings suggest that BBOWL provides interpretable measures of uncertainty quantification with little appreciable cost in predictive performance, supporting its use in practice.



\section{Discussion}
\label{s:discussion}

We have proposed backward Bayesian outcome weighted learning (BBOWL), a Bayesian adaptation of BOWL that combines sequential backward optimization with posterior uncertainty quantification. By embedding the Bayesian support vector machine structure into the recursive framework of BOWL, BBOWL propagates posterior uncertainty across stages of a dynamic treatment regime. Within this Bayesian formulation, the posterior distribution of the decision function naturally yields individual-level measures of uncertainty in treatment recommendations. This is an advantage not shared by many existing classification-based approaches to dynamic treatment regime estimation in multi-stage settings.

Across the simulated scenarios, BBOWL achieved classification performance similar to BOWL under varying sample sizes, stages, and prior specifications. Importantly, the ability to quantify decision uncertainty was obtained without a discernible loss in predictive accuracy. The small differences observed among the normal, exponential power, and spike-and-slab priors indicate that BBOWL's performance was relatively insensitive to the prior choice under the settings considered. In both terminal and intermediate outcome scenarios, BBOWL maintained low misclassification rates that decreased with increasing sample size, consistent with expected sample behavior in these simulation settings.

Several limitations warrant discussion. First, as in other outcome weighted learning approaches, the method relies on accurate specification of treatment assignment probabilities. Second, we restricted attention to binary treatments under randomized assignment with known propensity. Applications involving multi-arm, continuous dose, or nonrandomized exposures would require additional methodological development to address estimation of the propensity mechanism and potential confounding. Furthermore, additional preprocessing and sensitivity analyses may be necessary for applications to assess the impact of underlying missing data mechanisms \citep{Sun2025-mb}. Future work may also extend the research to accommodate censored outcomes, observational data, and adaptive trial designs \citep{Goldberg2012-im, Moodie2012-qb, Thall2007-hu}. Third, computational efficiency may be important in high dimensional settings. Although Gibbs sampling scales well in moderate dimensions, new computational strategies may be helpful in applications with many covariates.

Despite these limitations, BBOWL offers a novel framework for learning individualized treatment recommendations in multi-stage decision settings. Its ability to provide both point estimates and uncertainty measures makes it useful in clinical applications where interpretability is essential. BBOWL can readily be extended to reinforcement learning contexts such as health policy optimization, personalized education, and adaptive interventions in the social sciences.

Future research directions include formal assessment of posterior concentration for decision uncertainty, incorporation of nonlinear decision rules, and development of scalable algorithms for high-dimensional or large sample contexts. Nonetheless, we believe BBOWL offers a practical approach for estimating DTRs in multi-stage settings with the additional benefit of clinically relevant uncertainty measures associated with individual treatment decisions.


\backmatter






\bibliographystyle{biom} 
\bibliography{mybiblio}








\appendix

\label{lastpage}

\end{document}